\documentclass[preprint,showpacs,showkeys,preprintnumbers,amsmath,amssymb]{revtex4}
\usepackage{amssymb}
\usepackage{graphicx}
\usepackage{subfigure}
\usepackage{float}

\begin{document}

\title{A nonlocal variable coefficient modified KdV equation derived from two-layer fluid system and its exact solutions}

\author{ Xi-zhong Liu}
\affiliation{Institute of Nonlinear Science, Shaoxing University, Shaoxing 312000, China}
\begin{abstract}
A nonlocal form of a two-layer fluid system is proposed by a simple symmetry reduction, then by applying multiple scale method to it a general nonlocal two place variable coefficient modified KdV (VCmKdV) equation with shifted space and delayed time reversal is derived. Various exact solutions of the VCmKdV equation, including elliptic periodic waves, solitary waves and interaction solutions between solitons and periodic waves are obtained and analyzed graphically. As an illustration, an approximate solution of the original nonlocal two-layer fluid system is also given.
\end{abstract}



\maketitle

\section{Introduction}
In nonlinear science, to obtain exact solutions of nonlinear equations, including nonlinear wave solutions and soliton solutions, etc., is a basic and challenging task. In the past few decades, many effective methods have been developed to study a lot of integrable or non-integrable equations, such as Korteweg-de Vries (KdV) equation along with its multidimensional analog Kadomtsev-Petviashvili equation, different type of nonlinear Schr\"{o}dinger equations and so on, to give their various exact solutions. However, these well studied nonlinear systems are mostly local ones until Ablowitz and Musslimani \cite{muss} in 2013 introduced a PT symmetric nonlocal Schr\"{o}dinger (NNLS) equation
\begin{equation}\label{schr}
iq_{t}(x,t)=q_{xx}(x,t)\pm q(x,t)q^*(-x,t)q(x,t),
\end{equation}
with $\ast$ being complex conjugate and $q$ being a complex valued function of the real variables
$x$ and $t$. It is remarkable that, despite of the nonlocal property, Eq. \eqref{schr} is an integrable infinite dimensional Hamiltonian equation, which can be solved by inverse scattering transform and possesses infinitely number of conservation laws. Although Eq. \eqref{schr} was derived with physical intuition, since then, a wealth of new nonlocal nonlinear integrable infinite dimensional Hamiltonian dynamical systems are introduced from integrability requirement, among which, many of them are constructed by simple symmetry reductions of general AKNS scattering problems. These include reverse space-time, and in some cases reverse time, nonlocal nonlinear modified Korteweg-deVries equation, nonlocal sine-Gordon equation, derivative NNLS equation and so on \cite{markj}. At the same time, many efficient methods have been developed to obtain abundant wave solutions of nonlocal systems such as solitary waves, periodic waves, rogue waves, etc. \cite{fokas,maly,kh,zhangy}.

In fact, nonlocal phenomena exist commonly in many fields of real nature \cite{mau,coc,per,con}. To establish physically meaningful nonlocal systems, Lou proposed to construct two place Alice-Bob system (AB system) by using ``AB-BA equivalence principle'' and ``$\hat{P}_s$-$\hat{T}_d$-$C$ principle'' \cite{louab2}.  Through this way, many nonlocal version of physically important nonlinear systems are constructed, such as AB-KdV equation \cite{louab, jiaman}, AB-modified KdV equation \cite{licongcong}, AB-NLS equation \cite{louab2}, etc., and various exact solutions are given, including some types of shifted parity and time reversal ($\hat{P}_s\hat{T}_d$) symmetry breaking multi-soliton solutions, $\hat{P}_s\hat{T}_d$ conserving localized excitation, rogue wave solutions \cite{louqiao,yk} and so on. Especially, for some complicated physically important AB-type models, for example  AB-type multiple vortex interaction systems in atmosphere system, which are hard to give exact soliton solutions or periodic wave solutions, the multiple scale expansion method is applied to transform them into some aimed AB-type nonlinear equations, such as AB-(m)kdV equation \cite{tang,tangkdv}, AB-NLS equation \cite{tang2}, etc. Through this way, many approximate solutions for the original complicated AB-type systems can be derived from various exact solutions of aimed AB-type equations, such as periodic elliptic wave solutions, soliton solutions and the interaction solutions between solitons and periodic waves, etc., meanwhile, many physical phenomena like correlated dipole blocking event can be appropriately explained.

In this paper, we use a two-layer fluid system \cite{ped}
\begin{equation}\label{q1}
q_{1t}+J\{\psi_1,q_1\}+\beta\psi_{1x}=0,
\end{equation}
\begin{equation}\label{q2}
q_{2t}+J\{\psi_2,q_2\}+\beta\psi_{2x}=0,
\end{equation}
where
\begin{equation}\label{rq1}
q_1=\psi_{1xx}+\psi_{1yy}+F(\psi_2-\psi_1),
\end{equation}
\begin{equation}\label{rq2}
q_2=\psi_{2xx}+\psi_{2yy}+F(\psi_1-\psi_2),
\end{equation}
and $J\{a,b\}=a_xb_y-b_xa_y$, as a starting point to derive a nonlocal mKdV system and study its various solutions. In Eqs. \eqref{q1}-\eqref{rq2}, $F$ indicates coupling strength between two layers of fluid, which is a small constant; $\beta=\beta_0(L^2/U)$ where $\beta_0=(2\omega_0/a_0)\cos(\varphi_0)$, with $a_0$ being the earth's radius, $\omega_0$ being the angular frequency of the earth's rotation and $\varphi_0$ being the latitude, $U$ is the characteristic velocity scale while $L$ is characteristic horizontal length scale. In deriving Eqs. \eqref{q1}-\eqref{q2} in Ref. \cite{ped}, the constants are fixed as $L=10^6 m$ and $U=10^{-1} m s^{-1}$.

The paper is organized as follows. In Sect. \uppercase\expandafter{\romannumeral2},  a general nonlocal variable coefficient modified KdV (VCmKdV) equation with shifted parity and delayed time reversal is derived from nonlocal version of the two-layer system \eqref{q1}-\eqref{q2} by using multiple scale expansion method. In Sect. \uppercase\expandafter{\romannumeral3}, various exact solutions of the VCmKdV system are obtained and analyzed graphically, including elliptic wave solutions, solitary wave solutions, interaction solutions between elliptic waves and solitons. Especially, a kind of elliptic wave solution exhibits abundant wave structures for different choices of the coefficient functions of the VCmKdV system, which is analyzed and depicted by 4 cases. In Sect. \uppercase\expandafter{\romannumeral4}, as a simple illustration, an approximated solution to the original nonlocal version of the two-layer system is given by using a known solution of the VCmKdV system. The last section devotes to a summary and discussion.

\section{derivation of a nonlocal variable coefficient modified KdV equation }
 Based on the AB-BA equivalence principle and $\hat{P}_s$-$\hat{T}_d$-$C$ principle, the nonlocal counterpart of the two-layer system \eqref{q1}-\eqref{q2} can be obtained by employing the following symmetry constraint
\begin{equation}\label{pt1}
\psi_2=\hat{P}_s^x\hat{T}_d\psi_{1}=\psi_1(-x+x_0,y,-t+t_0),
\end{equation}
\begin{equation}\label{pt2}
q_2=\hat{P}_s^x\hat{T}_dq_{1}=q_1(-x+x_0,y,-t+t_0).
\end{equation}
To derive a nonlocal modified KdV type equation with shifted parity and delayed time reversal from it, according to the multiple scale expansion method, the following long wave approximation assumption is assumed
\begin{equation}\label{lvar}
\xi=\epsilon(x-c_0t), \, \tau=\epsilon^3t,
\end{equation}
where $\epsilon$ is a small parameter, and $c_0$ is an arbitrary constant. The stream function $\psi_1$ can be expanded as
\begin{equation}\label{epsi1}
\psi_1=c_0y+U_0+\psi_{11}(\xi,y,\tau),
\end{equation}
where $U_0\equiv U_0(y)$ is an arbitrary function of $y$ and the last term can be expanded as
\begin{equation}\label{epsi11}
\psi_{11}(\xi,y,\tau)=\epsilon\phi_{11}+\epsilon^2\phi_{12}+\epsilon^3\phi_{13}+O(\epsilon^4),
\end{equation}
with $\phi_{1i}\equiv\phi_{1,i}(\xi,y,\tau),\,(i=1,2,3)$ being functions of indicated variables. The model constants $F$ and $\beta$ can be reasonable assumed as $\epsilon$ and $\epsilon^2$ order, respectively, i.e.,
\begin{equation}\label{ec}
F=F_0\epsilon,\,\beta=\beta_1\epsilon^2,
\end{equation}
which means that coupling strength between two layers is small and the effect of the rotation of the earth is even smaller than that.

Substituting Eq. \eqref{epsi1} with Eqs. \eqref{lvar}, \eqref{epsi11} and \eqref{ec} into Eqs. \eqref{q1} and \eqref{q2} with \eqref{pt1}, \eqref{pt2} and vanishing coefficients of $O(\epsilon)$, we obtain
\begin{equation}\label{ep11}
(U_{0y}-2c_0)\phi_{11yy\xi}-U_{0yyy}\phi_{11\xi}=0,
\end{equation}
and
\begin{equation}\label{ep12}
(U_{0y}-2c_0)\phi_{21yy\xi}-U_{0yyy}\phi_{21\xi}=0,
\end{equation}
where $\phi_{21}=\hat{P}_s^{\xi}\hat{T}_d^{\tau}\phi_{11}=\phi_{11}(-\xi+\xi_0,y,-\tau+\tau_0)$.

Since Eqs. \eqref{ep11} and \eqref{ep12} are linear with respect to $\phi_{11}$ and $\phi_{12}$, we assume them in the variable separation form as
\begin{equation}\label{sophi11}
 \phi_{11} = G_{0}(y,\tau)A(\xi, \tau)\equiv G_{0}A,
\end{equation}
and
\begin{equation}\label{sophi21}
 \phi_{21} = P_{0}(y,\tau)B(\xi, \tau)\equiv P_{0}B,
\end{equation}
with $B(\xi,\tau)=\hat{P}_s^{\xi}\hat{T}_d^{\tau}A=A(-\xi+\xi0,-\tau+\tau_0)$ and $P_{0}(y,\tau)=\hat{T}_d^{\tau}G_{0}=G_{0}(y,-\tau+\tau_0)$. Now substitute Eqs. \eqref{sophi11} and \eqref{sophi21} into Eqs. \eqref{ep11} and \eqref{ep12} to get
\begin{equation}
(U_{0y}+2c_0)G_{0yy}-U_{0yyy}G_{0}=0,
\end{equation}
and
\begin{equation}
(U_{0y}+2c_0)P_{0yy}-U_{0yyy}P_{0}=0,
\end{equation}
which have a general form of solution for $G_{0}$ and $P_{0}$
\begin{equation}\label{g0}
G_{0}= (U_{0y}+2c_0)F_{1}(\tau)\int{\frac{1}{(U_{0y}+2c_0)^2}dy},
\end{equation}
and
\begin{equation}\label{g02}
P_{0}= (U_{0y}+2c_0)H_{1}(\tau)\int{\frac{1}{(U_{0y}+2c_0)^2}dy},
\end{equation}
with $H_{1}(\tau)=\hat{T}_d^{\tau}F_{1}(\tau)=F_{1}(-\tau+\tau_0)$ being an arbitrary function.

Now vanishing the coefficients of $O(\epsilon^2)$ leads to
\begin{equation}\label{eps21}
(U_{0y}+2c_0)\phi_{12yy\xi}-U_{0yyy}\phi_{12\xi}=
(F_0U_{0y}+2F_0c_0+\phi_{11yyy})\phi_{11\xi}
+F_0(U_{0y}+2c_0)\phi_{21\xi}-\phi_{11y}\phi_{11yy\xi},
\end{equation}
and
\begin{equation}\label{eps22}
(U_{0y}+2c_0)\phi_{22yy\xi}-U_{0yy}\phi_{22\xi}=(F_0U_{0y}+2F_0c_0+\phi_{11yyy})\phi_{21\xi}
+F_0(U_{0y}+2c_0)\phi_{11\xi}-\phi_{21y}\phi_{21yy\xi},
\end{equation}
where $\phi_{22}=\hat{P}_s^{\xi}\hat{T}_d^{\tau}\phi_{12}=\phi_{12}(-\xi+\xi_0,y,-\tau+\tau_0)$.

To solve the Eqs. \eqref{eps21} and \eqref{eps22}, it is readily verified that $\phi_{12}$ and $\phi_{22}$ can be taken in the form
\begin{equation}\label{exphi12}
\phi_{12} =(G_1B+G_2A+G_3A^2)G_{0}
\end{equation}
and
\begin{equation}\label{exphi22}
\phi_{22}=\hat{P}_s^{\xi}\hat{T}_d^{\tau}\phi_{12}=(P_1B+P_2A+P_3A^2)P_{0},
\end{equation}
where $P_i(y,\tau)=\hat{T}_d^{\tau}G_i(y,\tau)=G_i(y,-\tau+\tau_0)\, (i=1,\,2,\, 3)$, which  are determined by substituting Eqs. \eqref{exphi12}-\eqref{exphi22} into Eqs. \eqref{eps21}-\eqref{eps22} as
\begin{eqnarray}
G_1&=&-\frac{Q}{G_0}\bigg[F_8+\int\frac{F_7-\int F_0P_0Qdy}{Q^2}dy\bigg],\\
G_2&=&-\frac{Q}{G_0}\bigg[F_6+\int\frac{F_5-\int F_0G_0Qdy}{Q^2}dy\bigg],\\
G_3&=&-\frac{2Q}{G_0}\bigg[F_4+\int\frac{F_3-\int (G_0G_{0yyy}-G_{0yy}G_{0y})dy}{4Q^2}dy\bigg],
\end{eqnarray}
where $Q=U_{0y}+2c_0$ and $F_i,\,(i=3\cdots8)$ are arbitrary functions of $\tau$.

To take one step further by vanishing coefficients $O(\epsilon^3)$ we get
\begin{multline}\label{ep31}
(U_{0y}+2c_0)\phi_{13yy\xi}-U_{0yyy}\phi_{13\xi}= (F_0\phi_{21y}+\phi_{12yyy}+\beta_1)\phi_{11\xi}+(F_0U_{0y}+2F_0c_0+\phi_{11yyy})\phi_{12\xi}\\+F_0
(U_{0y}+2c_0)\phi_{22\xi}-(U_{0y}+2c_0)\phi_{11\xi\xi\xi}+(F_0\phi_{21\xi}-\phi_{12yy\xi})\phi_{11
y}-\phi_{12y}\phi_{11yy\xi}+\phi_{11yy\tau},
\end{multline}
\begin{multline}\label{ep32}
(U_{0y}+2c_0)\phi_{23yy\xi}-U_{0yyy}\phi_{23\xi}= (F_0\phi_{11y}+\phi_{22yyy}+\beta_1)\phi_{21\xi}+(F_0U_{0y}+2F_0c_0+\phi_{21yyy})\phi_{22\xi}\\+F_0
(U_{0y}+2c_0)\phi_{22\xi}-(U_{0y}+2c_0)\phi_{21\xi\xi\xi}+(F_0\phi_{11\xi}-\phi_{22yy\xi})\phi_{21
y}-\phi_{22y}\phi_{21yy\xi}+\phi_{21yy\tau},
\end{multline}
where  $\phi_{23}=\hat{P}_s^{\xi}\hat{T}_d^{\tau}\phi_{13}=\phi_{13}(-\xi+\xi_0,y,-\tau+\tau_0)$.

Without using $y-$average trick as in many literatures, we assume the general form of variable separation solutions of Eqs. \eqref{ep31} and \eqref{ep32} as
\begin{equation}\label{rphi13}
\phi_{13}= g_1\int A_{\xi}Bd\xi+g_2A^3+g_3A^2+g_4A+g_5AB+g_6B^2+g_7B+g_8A_{\xi\xi},
\end{equation}
\begin{equation}\label{rphi23}
\phi_{23}=\hat{P}_s^{\xi}\hat{T}_d^{\tau}\phi_{13}=p_1\int B_{\xi}Ad\xi+p_2B^3+p_3B^2+p_4B+p_5AB+p_6A^2+p_7A+p_8B_{\xi\xi},
\end{equation}
where $p_i(y,\tau)=\hat{T}_d^{\tau}g_i(y,\tau)=g_i(y,-\tau+\tau_0)\,(i=1\cdots8)$ are functions to be determined later. Substituting Eqs. \eqref{rphi13} and \eqref{rphi23} into Eqs. \eqref{ep31} and \eqref{ep32} and requiring coefficients of different derivatives of $A$ and $B$ be proportional to each other up to a constant level, we arrive at a general nonlocal modified KdV system:
\begin{equation}\label{nonkdvA}
A_{\tau}+(e_2 A^2+e_3 A+e_8 B+e_4)A_{\xi}+(e_5 A+e_6 B+e_7)B_{\xi}+e_1A_{\xi\xi\xi}+e_9A=0,
\end{equation}
\begin{equation}\label{nonkdvB}
B_{\tau}+(s_2 B^2+s_3 B+s_8 A+s_4)B_{\xi}+(s_5 B+s_6 A+s_7)A_{\xi}+s_1B_{\xi\xi\xi}+s_9B=0,
\end{equation}
where $e_9=\frac{F_{1\tau}}{F_1},e_i\equiv e_i(\tau)\,(i=1\cdots8)$ and $s_i(\tau)=\hat{T}_d^{\tau}e_i(\tau)=e_i(-\tau+\tau_0)\,(i=1\cdots9)$ are arbitrary functions, while $g_i\,(i=1\cdots8)$ in Eq. \eqref{rphi13} are determined by
\begin{eqnarray}
g_1&=&-Q(m_{16}+\int(m_{15}-\int2G_{1y}G_{0y}^2+(4G_0G_{1yy}-F_0P_0)G_{0y}+2G_0G_{0yy}
G_{1y}\nonumber\\&&+(e_5-e_8)G_{0yy}+F_0G_0P_{0y}+G_0^2G_{1yyy}dy)Q^{-2}dy),\nonumber\\
g_2&=&-3Q(m_{4}+\int\big(m_3-\int(G_0G_{3yy}-3G_{0yy}G_3)G_{0y}-4G_{3y}G_{0y}^2
+G_0^2G_{3yyy}\nonumber\\&&+2G_0G_{0yy}G_{3y}+3G_0G_3G_{0yyy}-e_2G_{0yy}dy\big)(3Q)^{-2}dy),\nonumber\\
g_3&=&-2Q(m_{6}+\int(m_{5}-\int2(G_0G_{2yy}-G_{0yy}G_2)G_{0y}-2G_{2y}G_{0y}^2+2F_0G_0U_{0y}
G_3\nonumber\\&&+4F_0G_0G_3c_0+G_0^2G_{2yyy}+2G_0G_{0yy}G_{2y}+2G_0G_2G_{0yyy}
-e_3G_{0yy}dy)(2Q)^{-2}dy),\nonumber\\
g_4&=&-Q(m_{8}+\int(m_{7}-\int F_0(G_2G_0+P_1P_0)U_{0y}+2F_0G_0G_2c_0+2F_0P_1P_0c_0
\nonumber\\&&+G_0\beta_1-e_4G_{0yy}dy)Q^{-2}dy),\nonumber\\
g_5&=&-Q(m_{10}+\int(m_9-\int(F_0P_0-G_0G_{1yy}-G_{0yy}G_1)G_{0y}-2G_{1y}G_{0y}^2+G_0G_1
G_{0yyy}\nonumber\\&&-e_5G_{0yy}dy)Q^{-2}dy),\nonumber\\
g_6&=&-2Q(m_{12}+\int(m_{11}-\int2F_0U_{0y}P_3P_0+4F_0P_3P_0c_0-e_6G_{0yy}dy)(2Q)^{-2}dy),\nonumber\\
g_7&=&-Q(m_{14}+\int(m_{13}-\int F_0(G_1G_0+P_2P_0)U_{0y}-e_7G_{0yy}+2F_0c_0(G_1G_0+P_2P_0)dy)Q^{-2}dy),\nonumber\\
g_8&=&-Q(m_{2}+\int(m_{1}+\int U_{0y}G_0+2G_0c_0+e_1G_{0yy}dy)Q^{-2}dy),\nonumber\\
\end{eqnarray}
with $m_i\equiv m_i(\tau)\, (i=1\cdots 16)$ being arbitrary functions.
\section{exact solutions of the nonlocal VCmKdV equation}

\subsection{exact solutions of the VCmKdV system with constant coefficients}

To give more interesting solutions of the VCmKdV equation \eqref{nonkdvA}, for simplicity, we first assume the coefficient of it are all constants. Under this assumption as well as $e_9=0$ in Eq. \eqref{nonkdvA}, i.e. $F_1(\tau)$ is a nonzero constant, we use elliptic function expansion method and generalized tanh expansion method to give some periodic wave solutions and interaction solutions between soliton and periodic waves, respectively.
\subsubsection{periodic wave solutions}
\begin{figure}
\centering
\subfigure[]{
\label{snab} 
\includegraphics[width=0.45\textwidth]{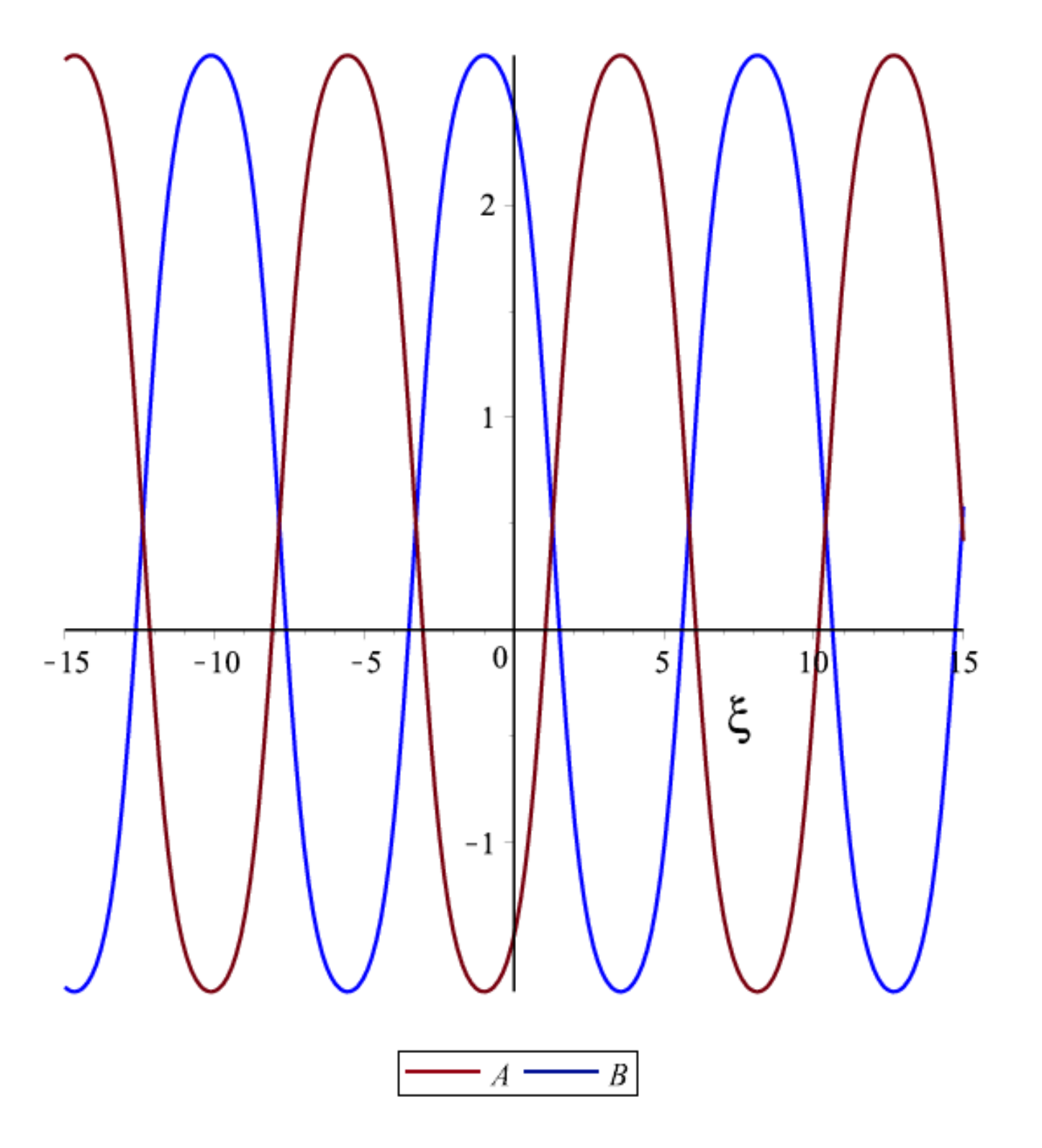}}
\subfigure[]{
\label{cna} 
\includegraphics[width=0.45\textwidth]{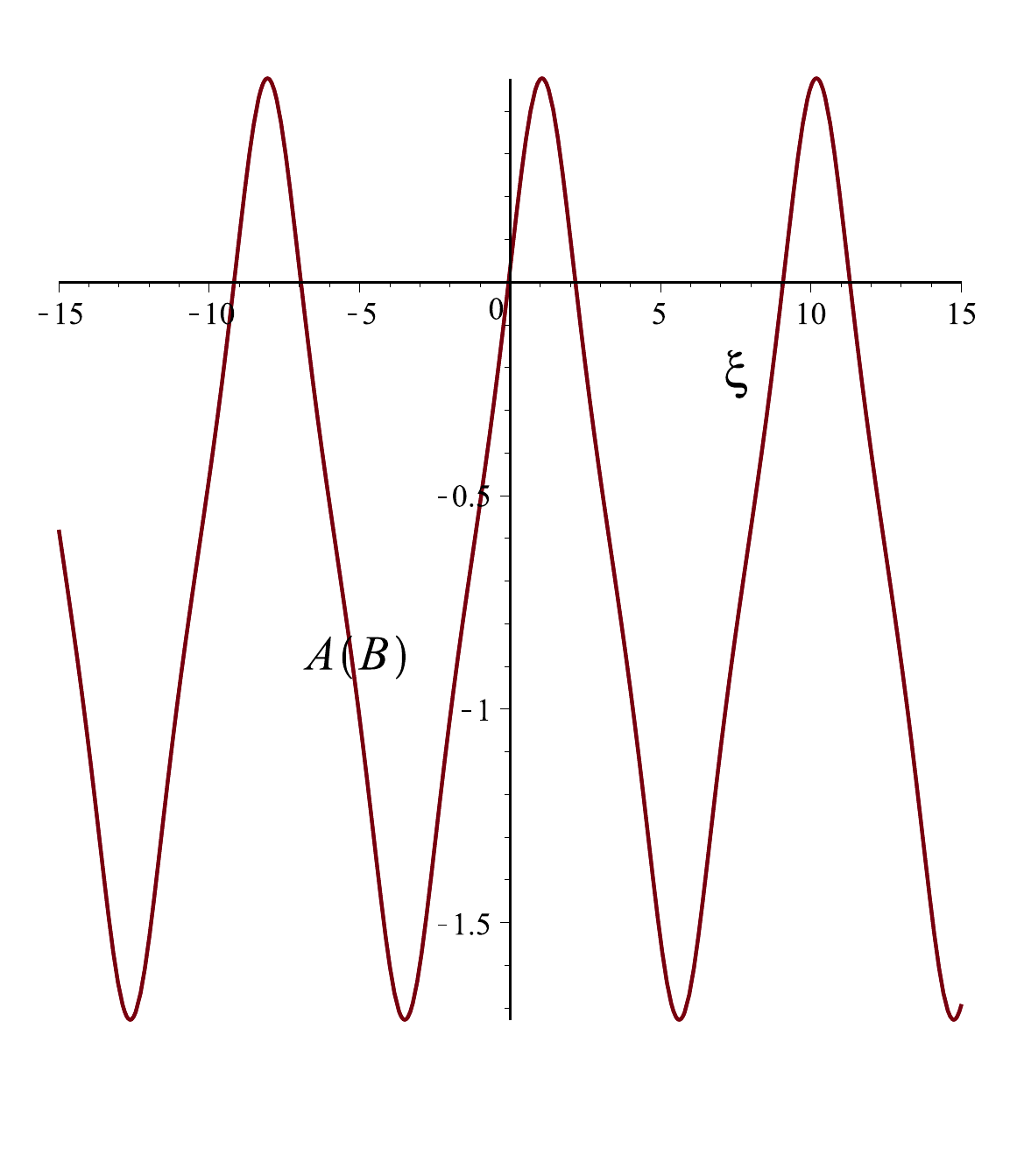}}
\caption{Profiles of elliptic periodic wave solution at a specific time $\tau=1$ with upper sign: (a) ${\rm sn}$ solution of \eqref{snsol} with \eqref{rb2}; (b) ${\rm cn}$ solution \eqref{cnsol} with \eqref{rb3}.}
\label{sncn} 
\end{figure}

\begin{figure}
\centering
\subfigure[]{
\label{snmab} 
\includegraphics[width=0.45\textwidth]{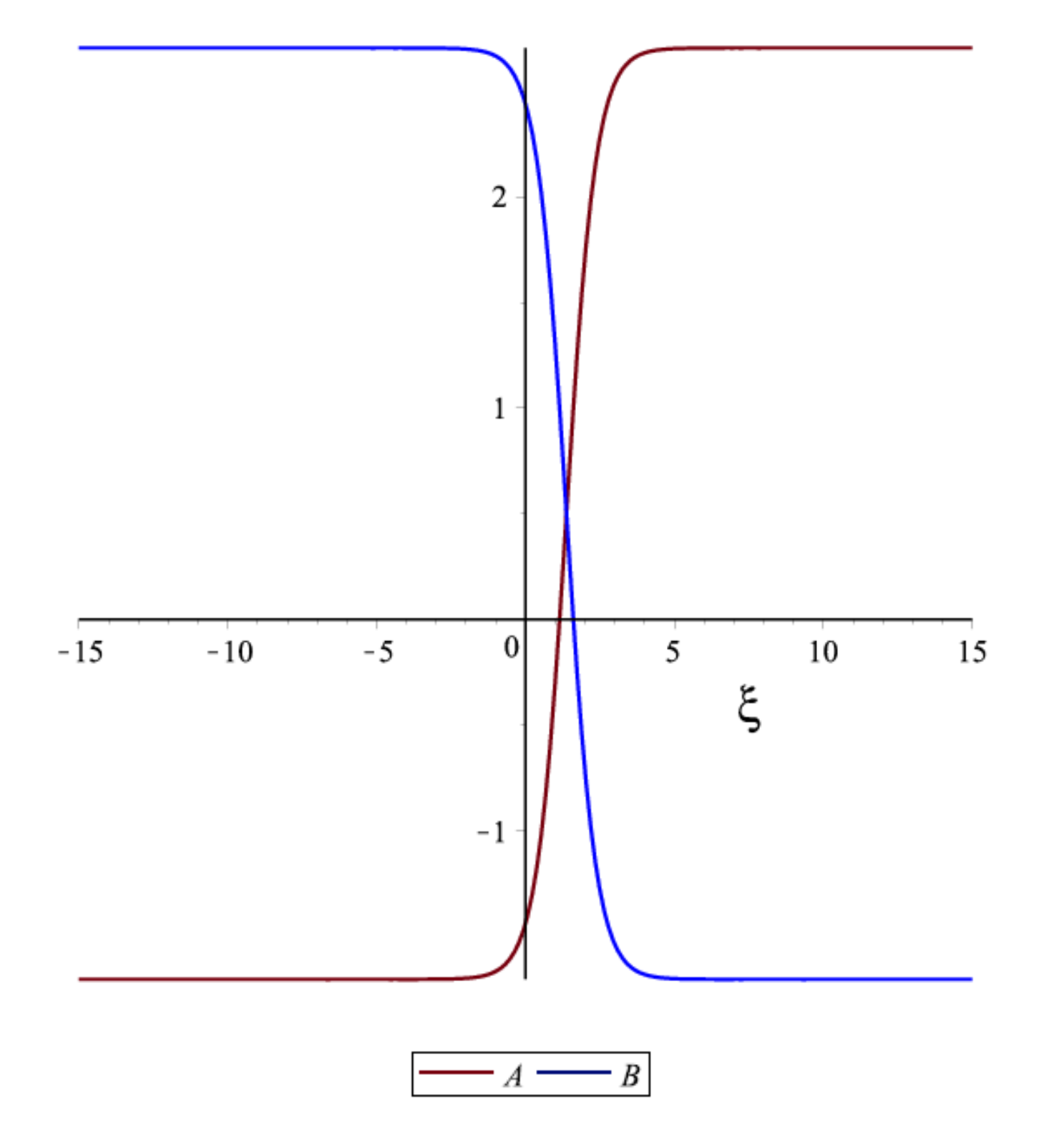}}
\subfigure[]{
\label{cnma} 
\includegraphics[width=0.45\textwidth]{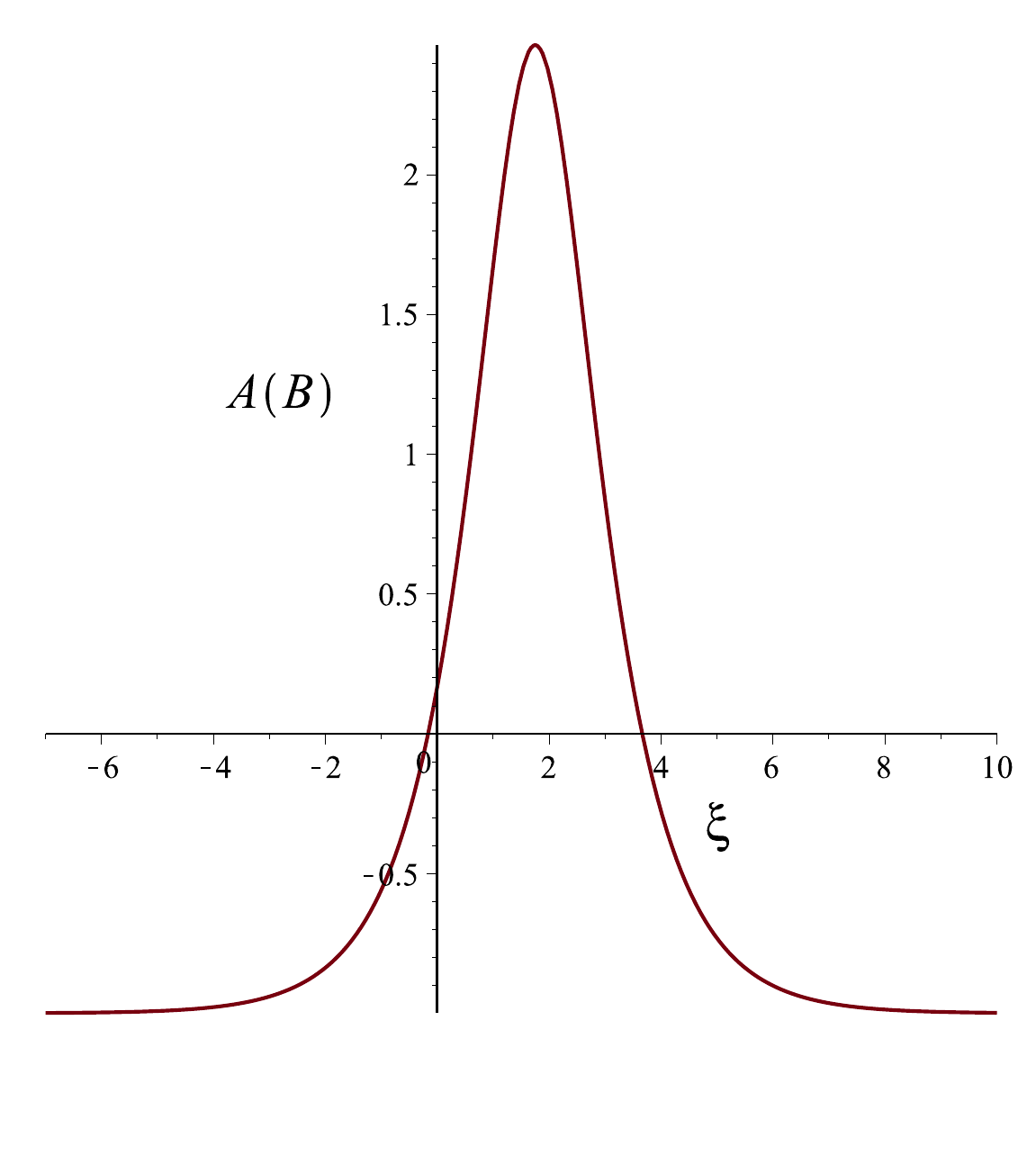}}
\caption{Profiles of soliton solution at a specific time $\tau=1$ with upper sign: (a) kink soliton solution of \eqref{snsol} with \eqref{rb2}; (b) bright soliton solution of \eqref{cnsol} with \eqref{rb3}.}
\label{inden} 
\end{figure}

Using elliptic function expansion method, after some routine work, it can be verified  Eq. \eqref{nonkdvA} admits two kinds of elliptic wave solutions, the first one is
\begin{equation}\label{snsol}
A=-\frac{e_3-e_5+e_6-e_8}{2e_2}\pm a_2m\sqrt{-\frac{6e_1}{e_2}} {\rm sn}\big(a_2(\xi-\frac{1}{2}\xi_0)+b_2(\tau-\frac{1}{2}\tau_0), m\big),
\end{equation}
with
\begin{equation}\label{rb2}
b_2=\frac{a_2\big[4e_1e_2(m^2+1)a_2^2-4(e_4-e_7)e_2
+(3e_8+e_3-3e_6-e_5)(e_3-e_5+e_6-e_8)\big]}{4e_2},
\end{equation}
and the other one is
\begin{equation}\label{cnsol}
A = -\frac{e_3+e_5+e_6+e_8}{2e_2}\pm a_3m\sqrt{\frac{6e_1}{e_2}}{\rm cn}\big(a_3(\xi-\frac{1}{2}\xi_0)+b_3(\tau-\frac{1}{2}\tau_0), m\big),
\end{equation}
with
\begin{equation}\label{rb3}
 b_3 =-\frac{a_3\big[4e_1(2m^2-1)e_2a_3^2+(4e_4+4e_7)e_2-(e_3+e_5+e_6+e_8)^2\big]}{4e_2},
\end{equation}
and the others being arbitrary constants. It's interesting to see that the solution of \eqref{snsol} is $\hat{P}_s^{\xi}\hat{T}_d^{\tau}$ symmetry breaking while  the solution of \eqref{cnsol} is $\hat{P}_s^{\xi}\hat{T}_d^{\tau}$ symmetry conserving, which are shown in Figs. \ref{snab} and \ref{cna}, respectively, with upper sign and the parameters are fixed as
\begin{equation}
a_2=e_2=e_3=e_4=e_6=e_7=e_8=\xi_0=\tau_0=1,e_1=-1,e_5=2,b_2=-1.56,m=0.9,
\end{equation}
for Fig. \ref{snab}, and
\begin{equation}
a_3=e_1=e_2=e_3=e_4=e_6=e_7=e_8=1,\xi_0=\tau_0=0,e_5=2,b_3=-1.06,m=0.9,
\end{equation}
for Fig. \ref{cna}, respectively, at a specific time $\tau=1$. When $m$ in Eqs. \eqref{snsol} and \eqref{cnsol} approaches to unity, it reduces to $\tanh$ and ${\rm sech}$ functions, respectively, which are shown in Figs. \ref{snmab} and \ref{cnma}, with upper sign and the parameters are fixed as
\begin{equation}
a_2=e_2=e_3=e_4=e_6=e_7=e_8=\xi_0=\tau_0=m=1,e_1=-1,e_5=2,b_2=-1.75,
\end{equation}
for Fig. \ref{snmab}, and
\begin{equation}
a_3=e_1=e_2=e_3=e_4=e_6=e_7=e_8=m=1,\xi_0=\tau_0=0,e_5=2,b_3=-1.44,
\end{equation}
for Fig. \ref{cnma}, respectively, at a specific time $\tau=1$.
\subsubsection{interaction solutions between solitons and periodic waves}
To obtain interaction solutions between solitons and periodic waves, a simple but effective way is using generalized $\tanh$ expansion method. To this end, we assume the solution of Eq. \eqref{nonkdvA} has the form $A=a_0+a_1\tanh(f)$ with $a_0,\,a_1$ and $f$ are all undetermined functions of $\xi$ and $\tau$. By carrying out the standard procedure, we have
\begin{equation}\label{tansol}
A=-\frac{\sqrt{-6e_1e_2}f_{\xi\xi}}{2 e_2 f_{\xi}}-\frac{e_3-e_5+e_6-e_8}{2e_2}+\frac{\sqrt{-6e_1e_2}f_{\xi}}{e_2}\tanh(f),
\end{equation}
with $f=\xi_1+h_1{\rm arctanh}(h_2{\rm sn}(h_3\xi_2, m))$, $\xi_1=k_1(\xi-\frac{1}{2}\xi_0)+\omega_1(\tau-\frac{1}{2}\tau_0)$,
$\xi_2=k_2(\xi-\frac{1}{2}\xi_0)+\omega_2(\tau-\frac{1}{2}\tau_0)$,  and the constants an be classified into 44 cases, here, we just list 2 of them:

case (1)
\begin{equation}
h_1=1,\,h_2=\pm\sqrt{m},\,h_3 =\frac{2k_1}{k_2(m+1)},
\end{equation}
\begin{eqnarray}
\omega_1&=&\frac{k_1}{4(m+1)^2e_2}\big\{[8k_1^2(m^2+14m+1)e_1-4(m+1)^2(e_4-e_7)]e_2
\nonumber\\&&+(m+1)^2(3e_8+e_3-3e_6-e_5)
(e_3-e_5+e_6-e_8)\big\},
\end{eqnarray}
\begin{eqnarray}
\omega_2&=&\frac{k_2}{4(m+1)^2e_2}\big\{(8k_1^2(5m^2+6m+5)e_1-4(m+1)^2(e_4-e_7))e_2
\nonumber\\&&+(m+1)^2(3e_8+e_3-3e_6-e_5)(e_3-e_5+e_6-e_8)\big\},
\end{eqnarray}

case (2)
\begin{equation}
h_1=\frac{1}{2},\,h_2=-m,\,h_3 =\frac{2k_1}{k_2m},
\end{equation}
\begin{eqnarray}
\omega_1&=&\frac{k_1}{4m^2e_2}\big\{[8k_1^2(m^2+3)e_1-4m^2(e_4-e_7)]e_2
\nonumber\\&&+m^2(3e_8+e_3-3e_6-e_5)(e_3-e_5+e_6-e_8)\big\},
\end{eqnarray}
\begin{eqnarray}
\omega_2&=&\frac{k_2}{4m^2e_2}\big\{[8k_1^2(5m^2-1)e_1-4m^2(e_4-e_7)]e_2
\nonumber\\&&+m^2(3e_8+e_3-3e_6-e_5)(e_3-e_5+e_6-e_8)\big\},
\end{eqnarray}
with the others remain free.

\begin{figure}
\centering
\subfigure[]{
\label{figsn} 
\includegraphics[width=0.45\textwidth]{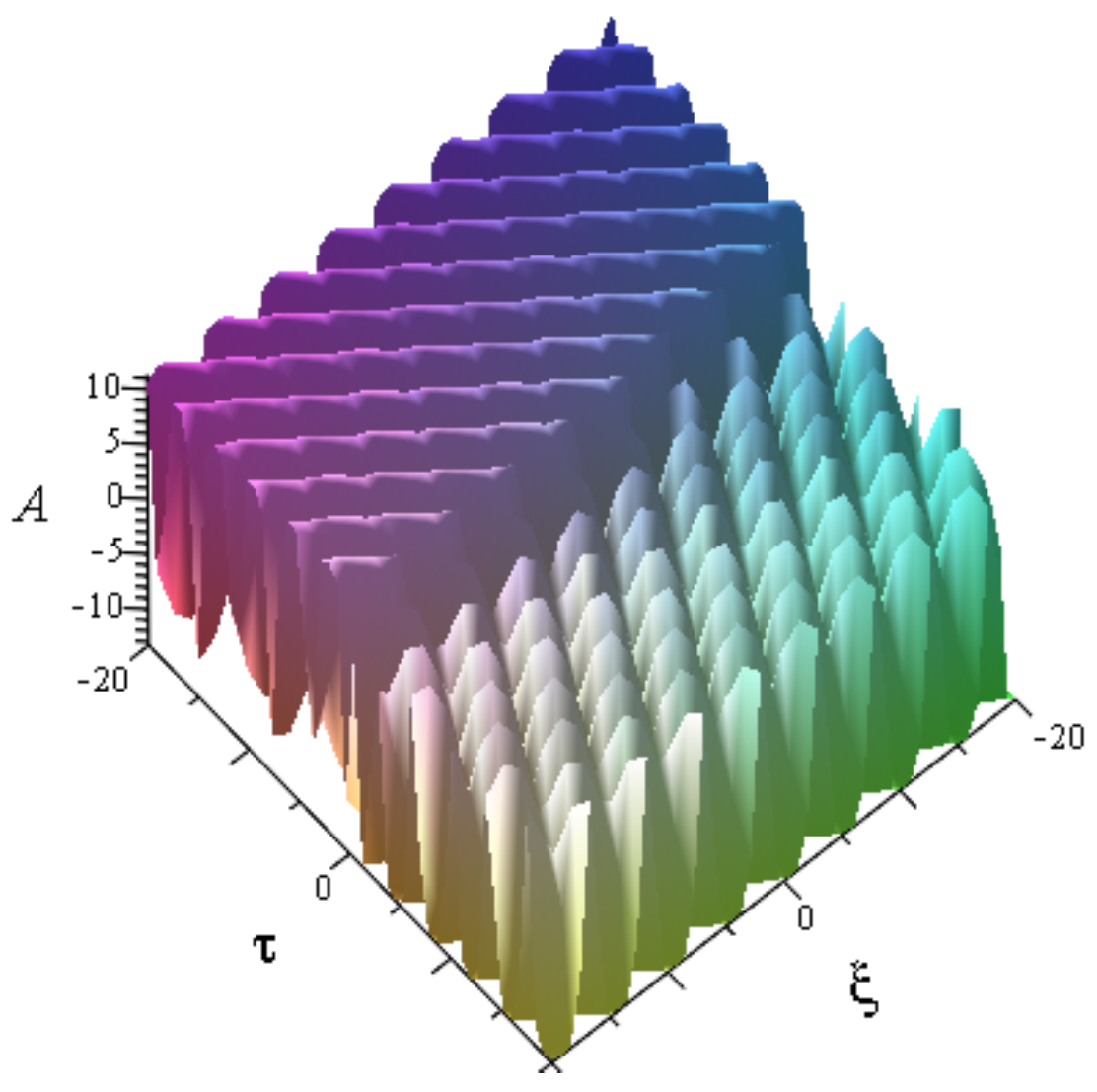}}
\subfigure[]{
\label{figcn} 
\includegraphics[width=0.45\textwidth]{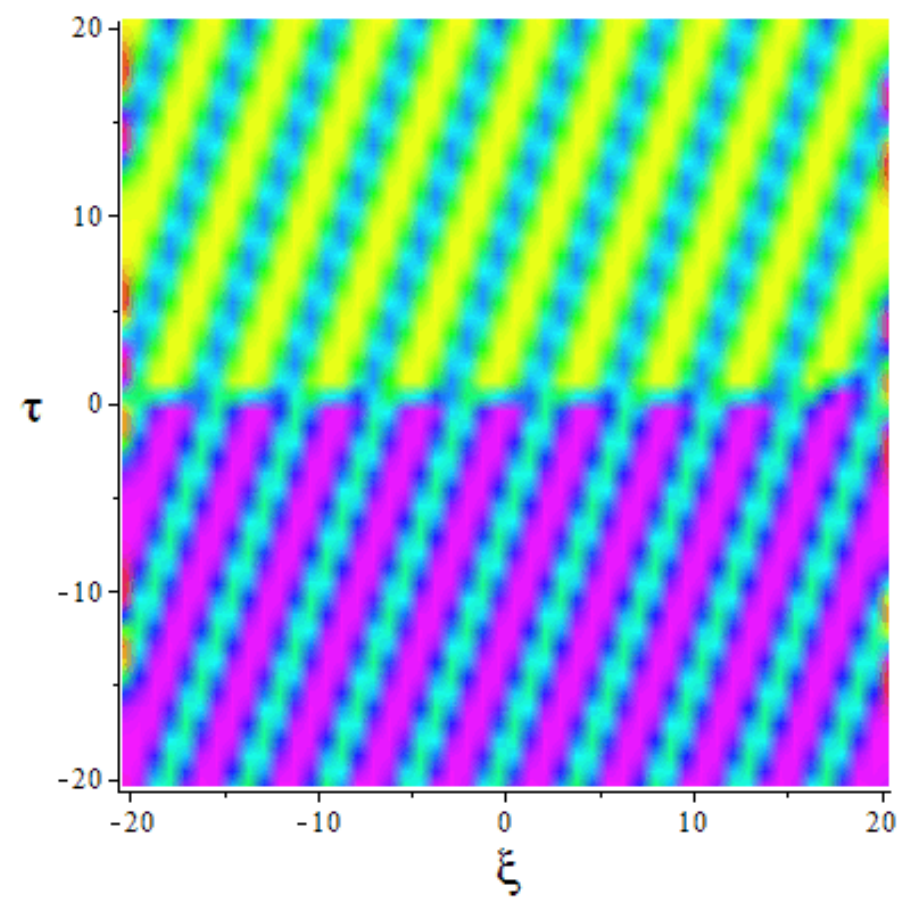}}
\caption{Profiles for $A$ of interaction solutions \eqref{tansol} in case (1): (a) three dimensional view; (b) density plot.}
\label{indenta} 
\end{figure}

\begin{figure}\label{inb}
\centering
\subfigure[]{
\label{ina} 
\includegraphics[width=0.45\textwidth]{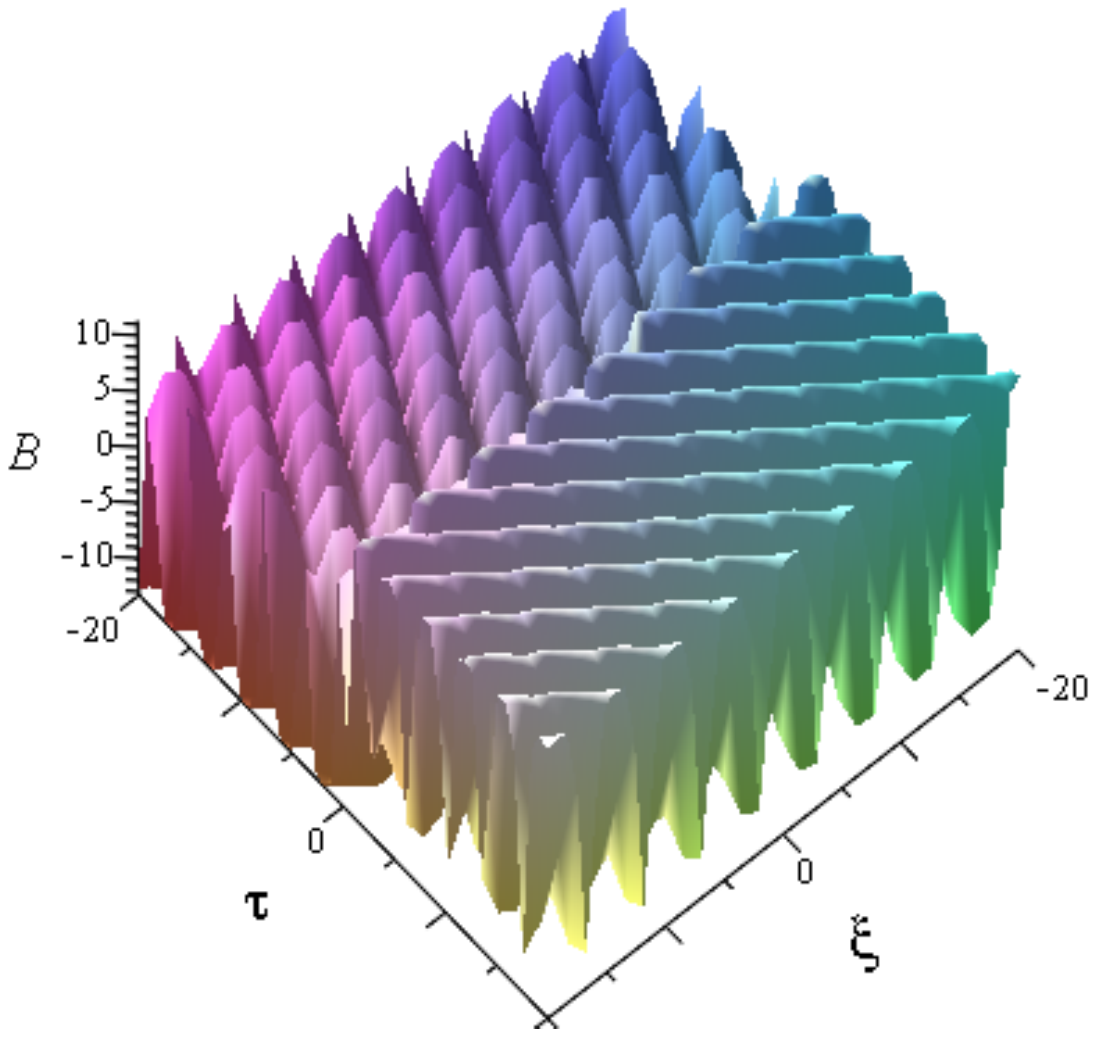}}
\subfigure[]{
\label{inb} 
\includegraphics[width=0.45\textwidth]{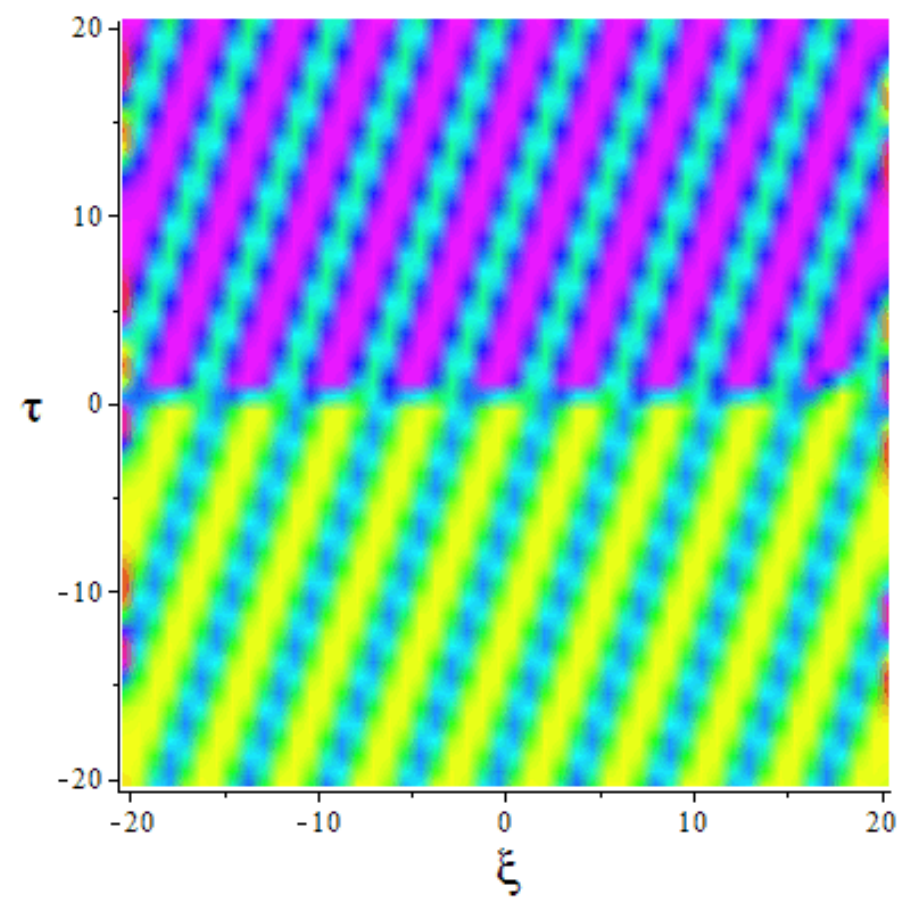}}
\caption{Profiles for $B=\hat{P}^{\xi}_s\hat{T}^{\tau}_dA$ of interaction solutions \eqref{tansol} in case (1): (a) three dimensional view; (b) density plot.}
\label{indentb} 
\end{figure}

To show the special feature of interaction solutions, case (1) of the solution \eqref{tansol} is displayed in Fig. \ref{indenta} for $A$ and  Fig. \ref{indentb} for $B$, with the parameters are fixed as
\begin{multline}
e_2=e_4=e_5=e_7=e_8=k_2=h_1=\xi_0=\tau_0=1,e_1=-0.1,e_3=2,e_6=2,k_1=8,\\m=0.6,h_2=0.77,h_3=10,\omega_1=-398.4,
\omega_2=-53.
\end{multline}
Fig. \ref{indenta} shows the structure of a kink soliton interacting with elliptic periodic waves, while Fig. \ref{indentb} shows the structure of a anti-kink soliton interacting with elliptic periodic waves, both of which are with a nonzero  phase change.

\subsection{exact solutions of the nonlocal VCmKdV equation with variable coefficients}
For the general case, i.e. $e_i(\tau)\,(i=1\cdots9)$ in Eq. \eqref{nonkdvA} are not constants, we give the following two kinds of periodic wave solutions, one is
\begin{equation}\label{cn2sol}
A =12k^2m^2{\rm cn^2}\big(k(\xi-\frac{1}{2}\xi_0)+\omega(\tau)(\tau-\frac{1}{2}\tau_0), m\big),
\end{equation}
with \begin{equation}
\omega(\tau)=\frac{-2}{2\tau-\tau_0}\int^{\tau-\frac{1}{2}\tau_0}k\big(4k^2(2m^2-1)e_1(\tau+\frac{1}{2}\tau_0)
+e_4(\tau+\frac{1}{2}\tau_0)+e_7(\tau+\frac{1}{2}\tau_0)\big)d\tau,
\end{equation}
and the other one is
\begin{equation}\label{sn2sol}
A =-12k^2m^2{\rm sn^2}\big(k(\xi-\frac{1}{2}\xi_0)+\omega(\tau)(\tau-\frac{1}{2}\tau_0), m\big),
\end{equation}
with
\begin{equation}
\omega(\tau)=\frac{2}{2\tau-\tau_0}\int^{\tau-\frac{1}{2}\tau_0}k\big(4k^2(m^2+1)e_1(\tau+\frac{1}{2}\tau_0)
-e_4(\tau+\frac{1}{2}\tau_0)-e_7(\tau+\frac{1}{2}\tau_0)\big)d\tau,
\end{equation}
and the others are arbitrary constants. In Eqs. \eqref{cn2sol} and \eqref{sn2sol}, {\rm sn} and {\rm cn} are elliptic sine function and cosine functions, respectively, with modulus $m$. In addition, both of these solutions are under the common condition
\begin{equation}
e_3(\tau) = e_1(\tau)-e_6(\tau)-e_5(\tau)-e_8(\tau),\, e_2(\tau)=e_9(\tau)=0.
\end{equation}

\begin{figure}
\centering
\subfigure[]{
\label{figsn} 
\includegraphics[width=0.45\textwidth]{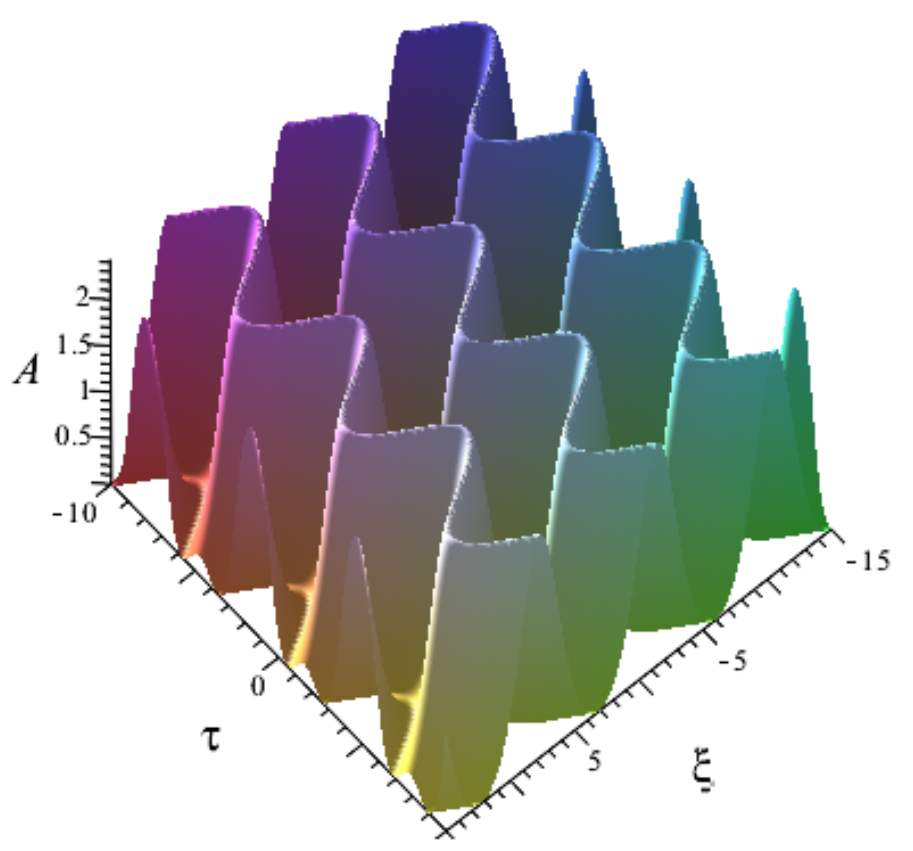}}
\subfigure[]{
\label{figcn} 
\includegraphics[width=0.45\textwidth]{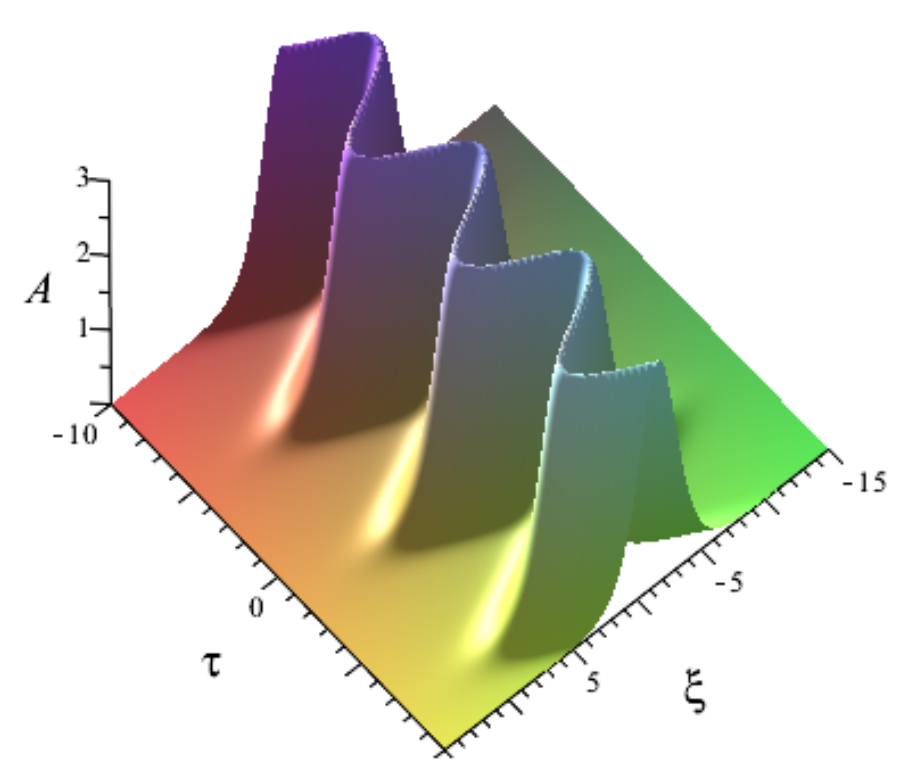}}
\caption{Profiles of the solution \eqref{cn2sol} with the parameters being fixed by \eqref{ex1} and (a)m=0.9; (b)m=1.}
\label{figex1} 
\end{figure}

\begin{figure}
\centering
\subfigure[]{
\label{figsn} 
\includegraphics[width=0.45\textwidth]{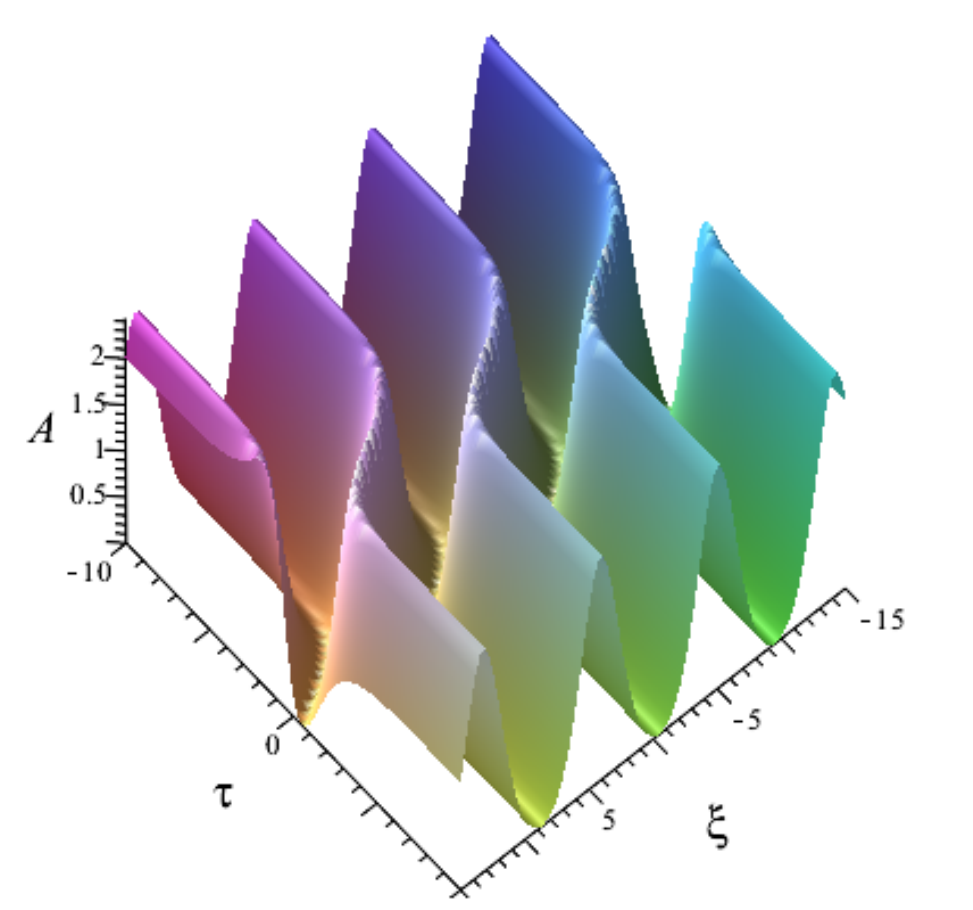}}
\subfigure[]{
\label{figcn} 
\includegraphics[width=0.45\textwidth]{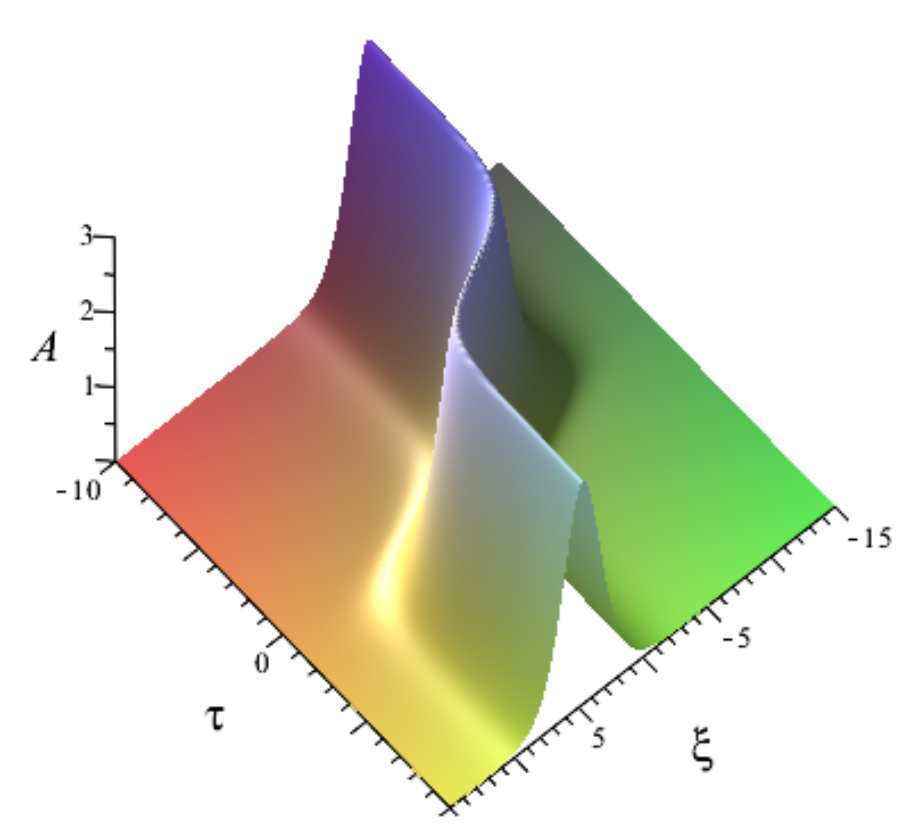}}
\caption{Profiles of the solution \eqref{cn2sol} with the parameters being fixed by \eqref{ex2} and (a)\emph{m}=0.9; (b)\emph{m}=1.}
\label{figex2} 
\end{figure}

\begin{figure}
\centering
\subfigure[]{
\label{figsn} 
\includegraphics[width=0.45\textwidth]{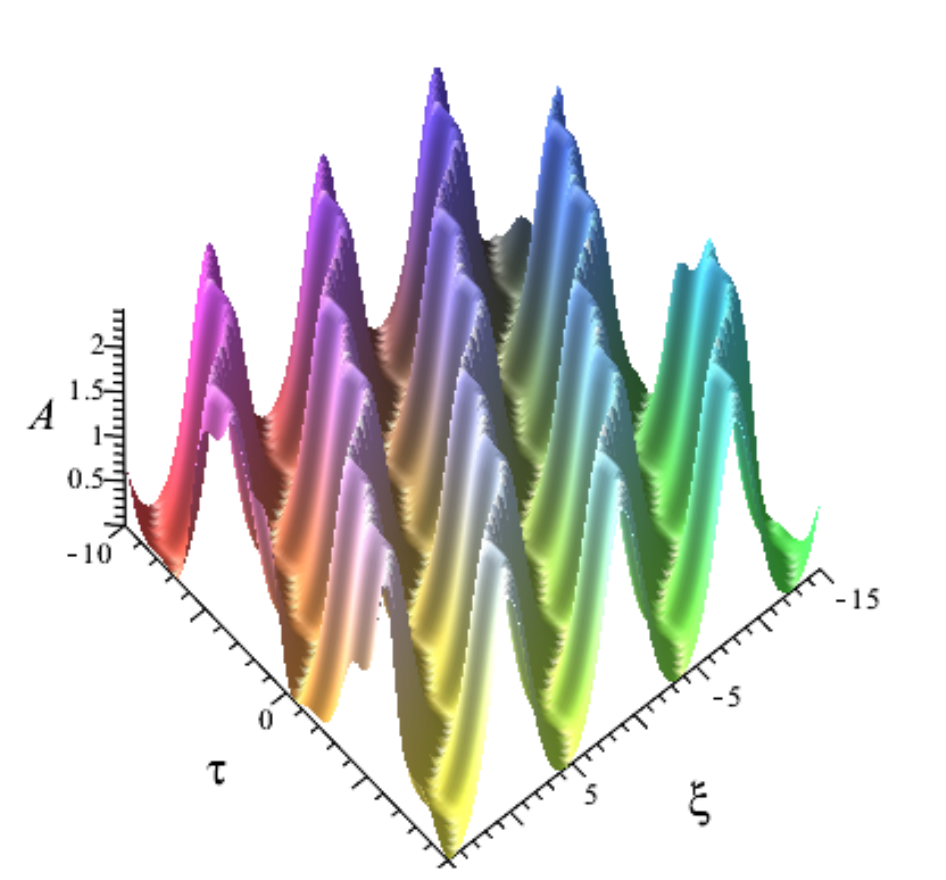}}
\subfigure[]{
\label{figcn} 
\includegraphics[width=0.45\textwidth]{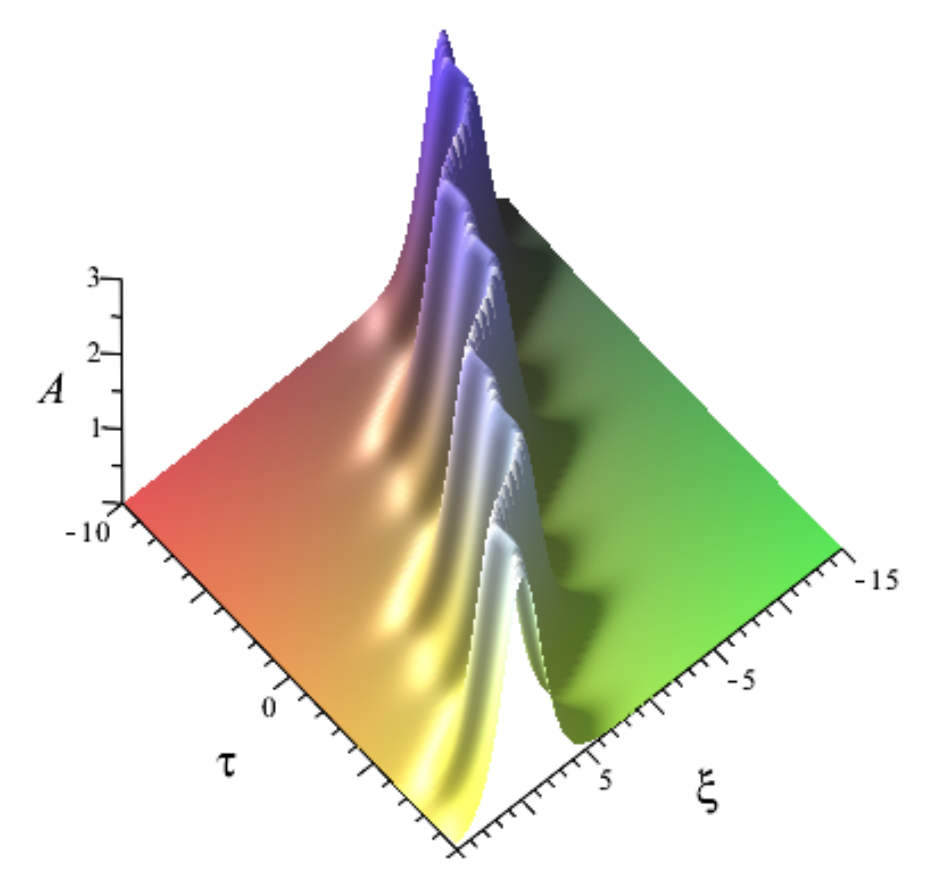}}
\caption{Profiles of the solution \eqref{cn2sol} with the parameters being fixed by \eqref{ex3} and (a)\emph{m}=0.9; (b)\emph{m}=1.}
\label{figex3} 
\end{figure}

\begin{figure}
\centering
\subfigure[]{
\label{figsn} 
\includegraphics[width=0.45\textwidth]{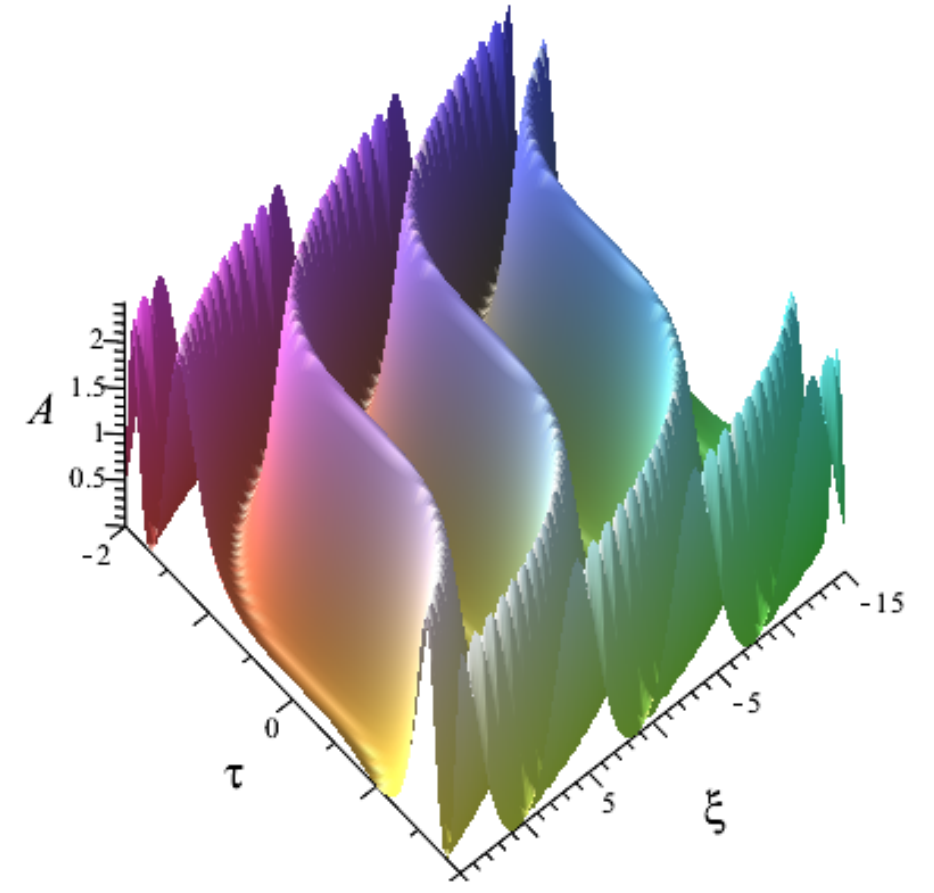}}
\subfigure[]{
\label{figcn} 
\includegraphics[width=0.45\textwidth]{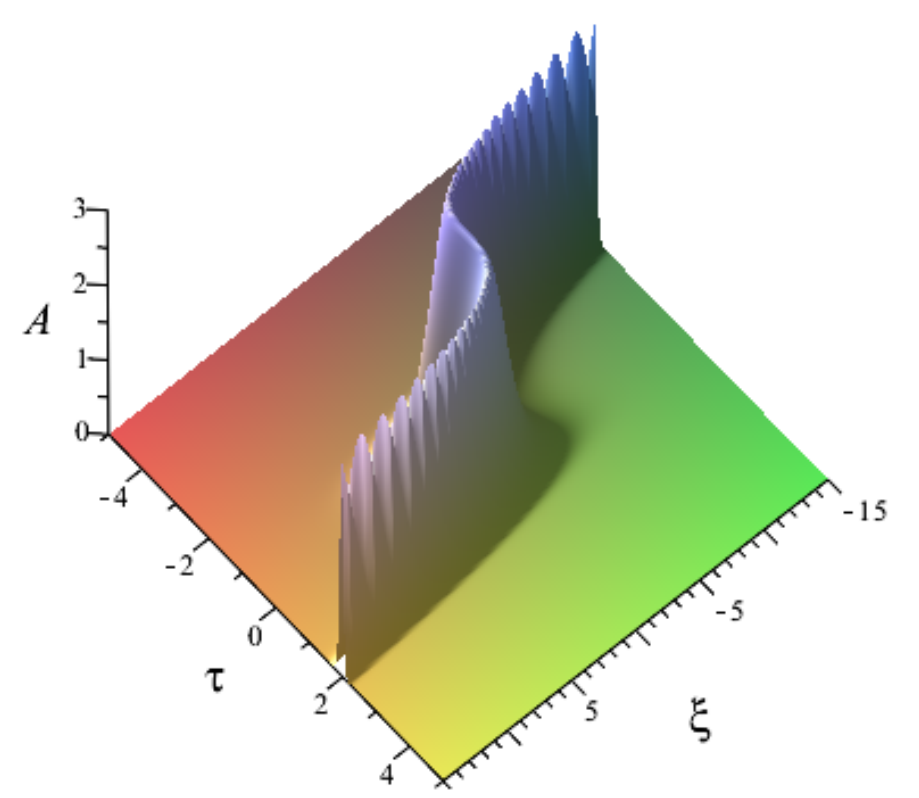}}
\caption{Profiles of the solution \eqref{cn2sol} with the parameters being fixed by \eqref{ex4} and (a)\emph{m}=0.9; (b)\emph{m}=1.}
\label{figex4} 
\end{figure}

Choosing different functions of $e_i(\tau),\,(i=1\cdots 8)$ results in rich structure of the solution \eqref{cn2sol}, here we show it graphically by fixing the parameters as
\begin{equation}\label{ex1}
e_1=e_4=e_5=e_6=e_7=e_8=\cos(\tau),e_3=-2\cos(\tau),k=0.5,\xi_0=\tau_0=0,
\end{equation}
for Fig. \ref{figex1},
\begin{equation}\label{ex2}
e_1=e_4=e_5=e_6=e_7=e_8={\rm sech}(\tau),e_3=-2{\rm sech}(\tau),k=0.5,\xi_0=\tau_0=0,
\end{equation}
for Fig. \ref{figex2},
\begin{equation}\label{ex3}
e_1=e_5=e_6=e_8=\cos(\tau),e_4=e_7=\cos(\tau)^2,e_3=-2\cos(\tau),k=0.5,\xi_0=\tau_0=0,
\end{equation}
for Fig. \ref{figex3},
\begin{equation}\label{ex4}
e_1=e_5=e_6=e_8=\tau^2,e_4=e_7=\tau^2\cosh(\tau),e_3=-2\tau^2,k=0.5,\xi_0=\tau_0=0,
\end{equation}
for Fig. \ref{figex4}, respectively. In each of the figures, the right wave is the limiting case of the left waves by taking $m\rightarrow1$ in Eq. \eqref{cn2sol}. It is interesting to observe from Figs. \ref{figex1}-\ref{figex4} that we can generate abundant wave shapes or wave dynamics by simply fixing different coefficients of the VCmKdV equation \eqref{nonkdvA}, which have potential applications in explaining real phenomenon of two-layer liquid system. In addition, it's clear that the solution \eqref{cn2sol} obviously conserve $\hat{P}_s^{\xi}\hat{T}_d^{\tau}$ symmetry, so the field $B$ have exactly the same dynamic behavior as $A$.

\section{An illustration: approximate solutions of the nonlocal two-layer fluid system}

\begin{figure}
\centering
\subfigure[]{
\label{strxy} 
\includegraphics[width=0.45\textwidth]{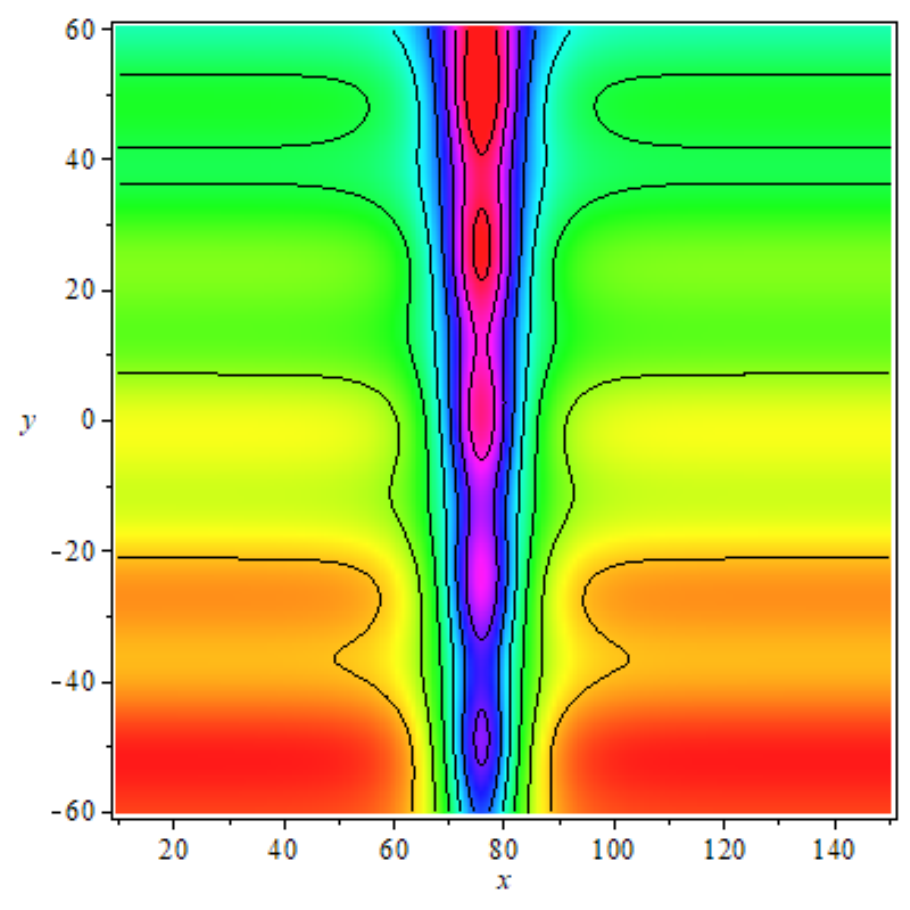}}
\subfigure[]{
\label{stryt} 
\includegraphics[width=0.45\textwidth]{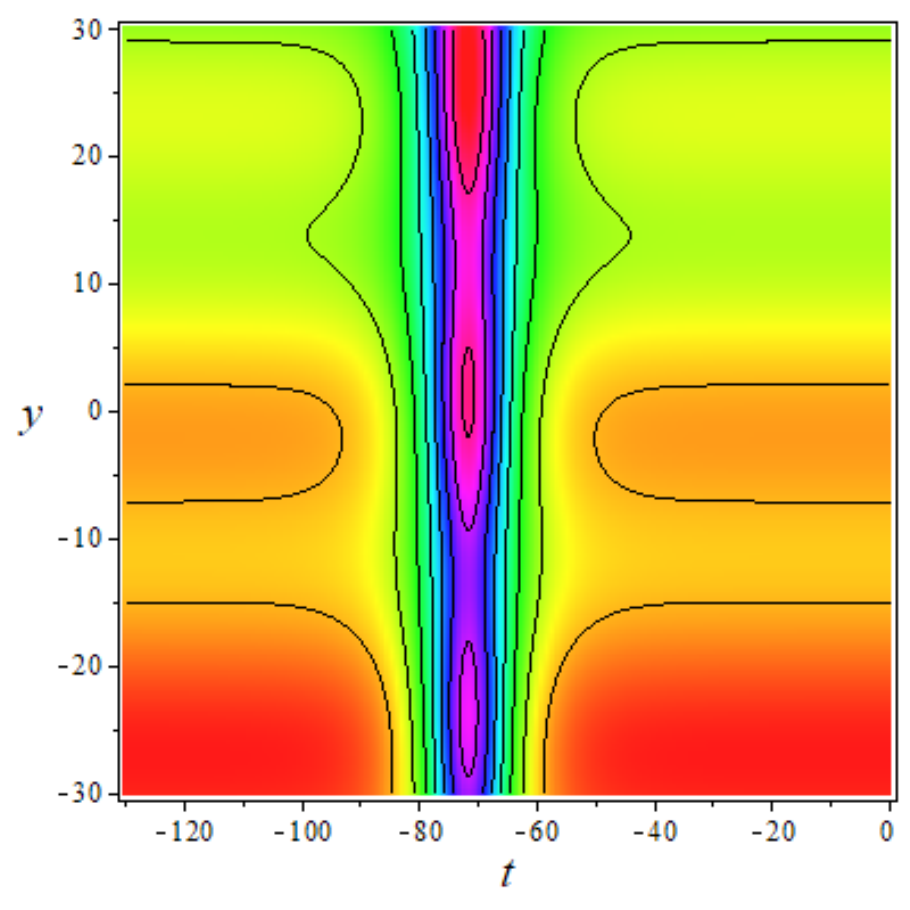}}
\caption{Density plot with stream lines of the approximate solution \eqref{aprpsi} with  \eqref{cnsol} at:(a)\emph{t}=1;(b)\emph{x}=1.}
\label{streamline} 
\end{figure}

Given the abundant known exact solutions of the nonlocal VCmKdV equation, it is meaningful to use them to generate approximate solutions of the original two-layer fluid system and give them appropriate physical explanations. To this end, up to $O(\epsilon)$, we first obtain the approximate solution of Eqs. \eqref{q1} and \eqref{q2} from Eqs. \eqref{epsi11} and \eqref{sophi11}
 \begin{equation}\label{aprpsi}
\psi_1=c_0y+U_0(y)+\epsilon G_0(y)A(\xi, \tau),\,\psi_{2}=\hat{P}_s^{\xi}\hat{T}_d^{\tau}\psi_{1}
\end{equation}
with $\xi=\epsilon(x-c_0t),  \tau=\epsilon^3t$, and $G_0$ is given by \eqref{g0}. When taking $A(\xi,\tau)$ in Eq. \eqref{aprpsi} as the solution of Eq. \eqref{cnsol}, i.e. the solution of \eqref{nonkdvA} with constant coefficients, and fixing the arbitrary functions and parameters as
\begin{equation}
U_0=\sin{\frac{y}{4}}
\end{equation}
and
\begin{multline}
C_1=0,\,C_2=200,\,a_3=2,\,b_3=-6,
e_1=e_2=e_3=e_5=e_6=e_7=e_8=c_0=1,\\e_4=2,\,m=1,\,t_0=10,\,x_0=160,\,\epsilon=0.1,
\end{multline}
and by substituting them into Eq. \eqref{aprpsi}, a special stream function is obtained, of which, the density distribution and streamlines are shown in Fig. \ref{streamline}. Observing from Figs. \ref{strxy} and \ref{stryt}, it's obvious that the vertices have the property of congestion both in space and time, which are mainly localized around the line of $x=75$ and $t=-71$, respectively. It should be noted that the existence of an arbitrary function $U_0$ in Eq. \eqref{aprpsi} could generate more kinds of approximate solutions for the original nonlocal two-layer fluid system. Furthermore, if we take the $A$ in Eq. \eqref{aprpsi} as the solution of Eq. \eqref{nonkdvA} with variable coefficients (e.g. the solution of \eqref{sn2sol}) more complex can be generated by choosing different $\omega(\tau)$, which have more potentiality of physical applications.
\section{Conclusion and discussion}
In summary, a nonlocal VCmKdV system with shifted parity and delayed time reversal is derived from a two-layer liquid system by applying AB-BA equivalence principle and multiple scale expansion method. Various exact solutions of the VCmKdV system are obtained, including elliptic periodic waves, solitary waves and interaction solutions between solitons and periodic waves. As an illustration, a simple approximate solution of the original nonlocal two-layer liquid system are given and analyzed.

From the derivation process of nonlocal VCmKdV system, it's clear that multiple scale method is powerful in obtaining aimed nonlocal nonlinear systems. In this sense, their exists a lot of work to derive other types of nonlocal equations from many physically important model and use their approximate solutions to analyze related phenomena, which needs to be explored in our future work.

\begin{acknowledgments}
 This work was supported by the National Natural Science Foundation of China under Grant Nos. 11405110 and the Natural Science Foundation of Zhejiang Province of China under Grant No. LY18A050001.
\end{acknowledgments}
\section*{Compliance with ethical standards}
\section*{Conflict of interest statement}
The authors declare that they have no conflicts of interest to this work. There is no professional or other personal interest of any nature or kind in any product that could be construed as influencing the position presented in the manuscript entitled.

\end{document}